\documentclass[a4paper,12pt]{article}

\usepackage[english]{babel}
\usepackage[utf8x]{inputenc}
\usepackage[T1]{fontenc}
\usepackage[hyphenbreaks]{breakurl}
\usepackage[hyphens]{url}

\usepackage[a4paper,top=2cm,bottom=2cm,left=3cm,right=3cm,marginparwidth=1.75cm]{geometry}

\usepackage{amsmath}
\usepackage{graphicx}
\usepackage[colorinlistoftodos]{todonotes}
\usepackage{hyperref}

\title{\vspace{-40pt}How Do App Stores Challenge the Global Internet Governance
Ecosystem?}
\author{
  Virgilio A. F. Almeida\\
  \texttt{Harvard University}
  \and
  Danilo Doneda\\
  \texttt{Rio de Janeiro State University}
  \and
  Carolina Rossini\\
  \texttt{Public Knowledge}
}
\date{}

\begin{document}
\maketitle

\begin{verbatim}
App stores challenge the culture of openness and resistance to central
authorities cultivated by the pioneers of the Internet. Could
multistakeholder governance bodies bring more inclusivity into the
global cyberspace governance ecosystem?
\end{verbatim}

Over four decades, a variety of stakeholders have been working both
formally and informally to build and maintain the governance ecosystem
that supports a unique and free Internet~{[}1{]}. To allow the Internet
to realize its full potential, the organizations involved in its
governance aim to keep the Internet open, secure, trustworthy, and
accessible to all~{[}2{]}. In the 90s Tim Berners-Lee introduced the
World Wide Web~{[}3{]}, with many new ideas that focused on a
decentralized and open Internet, where no permission was needed from a
central authority to post anything and with no central controlling
nodes. However, in the last few years, there has been a shift in the way
users access information and services on the Internet. Users have been
migrating from general-purpose Web browsers to singlepurpose apps, which
are standardized pieces of software designed to run on specific devices
and written for a specific operating system. There are more than 1
million apps available in the market today, that have been downloaded
more than 100 billion times~{[}4{]}.

The surge of apps for smartphones has brought innovation, new
technologies, and new entrepreneurs into the digital economy, but it
also introduced a new set of control points, represented by the app
stores~{[}5{]}. They're online digital distribution platforms from which
apps can be downloaded. App stores are associated with a specific
operating systems, and the most popular platforms are Android (Google),
iOS (Apple), and Windows Apps (Microsoft). The app stores challenge the
culture of openness and resistance to central authorities cultivated by
the pioneers of the Internet. Several characteristics of the app stores
emphasize the notion of new control points in the Internet governance
process, such as the submission processes, the levels of certification
and quality control required to have the app accepted, pricing
mechanisms, and criteria used to rank and suggest apps to users.

Multistakeholder support, security, privacy, transparency, neutrality,
freedom of expression, and competition are central values in the global
Internet governance ecosystem, where many players with different legal
statuses operate on a variety of layers --- on local, national,
regional, and international levels --- driven by technical innovation,
user needs, market opportunities, and political interests. So, a natural
question that arises is: How do app stores fit in the global Internet
governance process?

\section*{The Market's Size}\label{the-markets-size}

The importance of apps market and its potential ability to shape
innovation, competition, and consumer choice and behavior become more
vivid when compared with other core sources of access to communication
and content platforms. Since 2014, globally, consumers are connecting
more through mobile devices than desktop. In the US, for instance,
combined with mobile Web, mobile usage accounts for 60 percent of time
spent in digital media, while desktop-based digital media consumption
makes up the remaining 40 percent~{[}6{]}.

Not only has the method of connection changed, but the time spent with
these devices has changed as well. Last year in the US, for every 5.6
hours of time that consumers spent per day with digital media, it was
split with 2.8 hours going to mobile devices.

\section*{Policy Issues}\label{policy-issues}

There are a number of issues related to app stores that should be
followed by policy makers. For instance, the app store market model had
a major impact on transparency and openness, due to the fact that mobile
Internet users are turning away from browsers and are relying on apps.
Apps themselves don't depend on a browser, which makes less relevant
some user-empowering characteristics such as browser controls and Web
standards. In turn, the operating system's interface and its design
options make operating systems regain much of their former importance,
because available options and predefined standards (as well as
non-negotiable OS characteristics) are entirely operation
system-dependent. And it's not a coincidence that operating systems are
often designed by the same entities that controls app store
markets~{[}7{]}.

This situation can be regarded as a setback from the decentralized and,
as was then regarded, ever-growing Web. Not that open Web standards
don't play a most important role anymore --- for several reasons, and
not the least because even if several services are basically accessed
via apps, many of them are part of a complex platform that were built
upon or rely on a Web interface. But the empowerment of a few players
that run app store markets in detriment of a decentralized architecture
is evident, and it poses several questions regarding not only
competition issues, but as more of our interest at this time, user's
rights. This system change represents a shift of power from Internet
governance bodies to the companies responsible for the operating system.

Additionally, the submission process controlled by the owners of the app
stores transform them into arbiters of the freedom of expression and
right to innovate, for apps are themselves a conduit of ideas and
messages. The lack of transparency regarding how apps are ranked and
suggested to users and buyers in the main stores is another aspect that
shows the importance of discussing the role of app stores in the global
Internet governance ecosystem.

Another relevant issue has to do with freedom of expression values. One
consequence of a closed-app store ecosystem is that, when content is one
of the criteria used by app store managers to approve an app, the
judgment regarding content evaluation might harm freedom of expression.
In an example of such an episode, Frank Pasquale mentions that one app
store --- which has strict guidelines forbidding apps with
``objectionable content'' --- rejected an app designed for downloading
and formatting public domain texts on the basis of the possibility it
could be used to access a version of Kama Sutra~{[}9{]}.

Related to privacy, some specific characteristics of the app store
market model call attention. First, users' identities are much more
likely to be tied to the use of an app than is foreseeable on the Web,
due to the app store's control over the download and installation
process for a particular device and the constant pressure for
integration of data from one app to another app or Web-based service
account. This, added to the fact that app store managers generally are
in possession of users' billing data, makes the set of users' data they
handle potentially much more robust than on an open environment, making
it also much more difficult to use apps anonymously, for example ---
which is different from the Web.

The importance of the operating system is also visible when privacy
controls are considered. Although apps can have their own setting panels
where users can shape their privacy preferences, they aren't
standardized or present in every app. This makes for a concrete
simplification in the management of privacy settings. On the other hand,
the OS settings panels have developed increasingly important privacy
options, either global or app-specific, where users' options on issues
such as disclosure of geo-locational data, access to hardware (the
microphone, camera, and so on), access to stored data (such as
contacts), ad options, and others are available. Other instruments as
the (sometimes mandatory) display of information related to the user's
privacy during an app's normal use (such as a reminder that locational
data is being collected) can make for a more robust privacy environment.
This can't be taken for granted, as the privacy environment is, after
all, controlled by the OS manufacturer, which can shape the setting and
the available options based on several factors and not merely on the
presumed user's interest.

Depending on the OS management, app standards can be tuned to increase
users' privacy. As an example, health apps have recently shown a move
for more transparency and control in collecting and processing users'
data, which some operating systems are requiring from app developers.
This is a clear indication that the use of personal data is considered
particularly relevant or sensitive (such as locational data), given the
need for the user's explicit consent for gathering such data or the
mandatory demand for renewed consent in some situations~{[}10{]}.

Additionally, as apps grow more and more popular, domain names will lose
their importance, because Internet traffic mainly will go to mobile apps
and not to websites in a browser. ICANN continues to control the domain
name system, but app stores will control the mechanisms that lead users
to information and services~{[}11{]}.

For all the negative aspects of the closed-app ecosystems, they might
offer certain advantages for regulated areas such as clinical health
study. The control point established by the stores raises other
governance questions. On the positive side, more control can improve
security by making it harder for hackers and criminals to insert
malicious software in apps. On the other side, the app-submission
process is viewed by developers as restrictive. A recent survey~{[}8{]}
shows that a significant percentage of developers think stores and app
publishers use anticompetitive practices.

Also, the recent entry of big technology enterprises into clinical
health study in the US offers early evidence of how app ecosystems,
which are generally unregulated by technology policy, ironically might
assist in the enforcement of federal policy in other spaces. Clinical
studies in the US fall under an umbrella of rules and regulations
designed to protect the human ``subjects'' from unethical
experimentation or from harm. These protections extend into app-based
studies. Also, in terms of implementing privacy-friendly or pro-consumer
policies, app stores can require independent ethical review and informed
consent processes, a practice that at least a large competitor uses. In
this instance, the more-controlled system holds the possibility of
stricter enforcement of privacy and ethics.

\section*{}
\vspace{-30pt}
Multistakeholder governance mechanisms engaging diverse stakeholders
have proven that they can keep the Internet open, transparent, and
inclusive. To deal with critical issues such as freedom of expression,
ethics, privacy, and data protection, multistakeholder governance bodies
could be a promising alternative to support and address the limits of
app stores, while also bringing these players closer to the global
cyberspace governance ecosystem.

\section*{Acknowledgment}\label{acknowledgment}

We thank Yasodara Córdova for her valuable insights and suggestions.

\section*{References}\label{references}

\begin{enumerate}
\begin{small}
\item V.G. Cerf, ``The Governance Ecosystem, Comm. ACM, vol.~57, no. 4, 2014, p.~7.
\item The Global Commission on Internet Governance, One Internet, policy report, Centre for International Governance Innovation, Chatham House, and the Royal Inst. of Int'l Affairs, 2016; \url{http://www.chathamhouse.org/about/structure/international-security-department/global-commission-internet-governance-project}.
\item T. Berners-Lee et al., ``The World-Wide Web, Comm. ACM, vol.~37, no. 8, 1994, pp.~76--82.
\item M. Kende, The U.S. Mobile App Report, 2015, tech.report, comScore, 2016; \url{http://www.comscore.com/Insights/Presentations-and-Whitepapers/2015/The-2015-US-Mobile-App-Report}.
\item Y. Benkler, ``Degrees of Freedom, Dimensions of Power,'' Daedalus, vol.~145, no. 1, 2016, pp.~18--32.
\item S. Perez, ``Majority of Digital Media Consumption Now Takes Place in Mobile Apps,'' Tech Crunch, 21 Aug. 2014; \url{https://techcrunch.com/2014/08/21/majority-of-digital-media-consumption-now-takes-place-in-mobile-apps}.
\item S. Zuboff, ``Big Other: Surveillance Capitalism and the Prospects of Information Civilization,'' J. Information Technology, vol. 30, no. 1, 2015, pp.~775--789.
\item K. Turner, ``Developers Consider Apple's App Store Restrictive and Anticompetitive, Report Shows,'' Washington Post, 15 July 2016; \url{http://wpo.st/LMmx1}.
\item Frank Pasquale, The Black Box Society: The Secret Algorithms That Control Money and Information, Harvard Univ. Press, 2015.
\item Information, Harvard Univ. Press, 2015. S. Crawford, J. Zittrain, and L. Brem, Game Changers: Mobile Gaming Apps and Data Privacy, background note APSW04, Harvard Law School, 2012.
\item I. Burrington, ``Who Controls the Internet? Ted Cruz's Fantasy vs.~the Reality,'' Fusion, 6 Sept. 2016; \url{http://fusion.net/story/343533/who-controls-the-internet}.
\end{small}
\end{enumerate}

\end{document}